\documentclass[11pt,preprint,showpacs]{revtex4}
\usepackage{graphicx}
\usepackage{dcolumn}
\usepackage{bm}
\usepackage{amsmath}
\usepackage{amssymb}
\usepackage{amsfonts}

\begin{document}

\title{A new form of three-body Faddeev equations in the continuum}

\author{H.~Wita{\l}a$^1$}

\author{W.\ Gl\"ockle$^2$}

\affiliation{$^1$ M. Smoluchowski Institute of Physics, Jagiellonian
 University, PL-30059 Krak\'ow, Poland}

\affiliation{$^2$ Institut f\"ur theoretische Physik II,
Ruhr-Universit\"at Bochum, D-44780 Bochum, Germany}

\date{\today}

\begin{abstract}
We propose a novel approach to solve the  three-nucleon (3N) 
Faddeev equation which 
 avoids the  complicated singularity pattern going with the  moving 
logarithmic singularities of the standard approach. In this new 
approach the treatment of the 3N Faddeev equation becomes 
essentially as simple 
as the treatment of the two-body Lippmann-Schwinger equation. 
Very good agreement  of the new and old approaches in the application to 
nucleon-deuteron elastic scattering and the breakup reaction is found.
\end{abstract}


\maketitle

\section{Introduction}

The three-body Faddeev equations~\cite{Faddeev} in the continuum put three-body
  scattering on a firm mathematical basis. The complicated
  asymptotic boundary conditions in configuration space~\cite{glo83} are fully
  included. At the time of their formulation the computer power, however,
  was insufficient to solve them directly given local two-body
  forces. Therefore one started with forces of finite rank (the
  simplest being a  rank one separable force), which turns the
  three-body problem in a partial wave basis into a finite set
  of coupled integral equations in one variable - and thus
  feasible at that time. Nevertheless the way it was formulated in
  momentum space the free three-body propagator lead to a complicated
singularity structure of
  the integral kernel, namely to logarithmic singularities, whose position
  depend on the external variable - so called moving
  singularities~\cite{Schmid-Ziegelmann}. In order to avoid
this obstacle the path of
  integration in the kernel is moved into the complex plane (contour
  deformation), which imposes of course conditions on the analytical
  properties of the two-body separable forces. The need, however, for using
  the realistic two-body forces, which are dominantly local,
  enforced their expansion into a series of finite rank forces~\cite{hiden86},
  which was tedious and finally overcome by integrating the
  logarithmic singularities directly on 
the real momentum axis~\cite{kloet75}. It
  took until the eighties~\cite{wit86} that fully realistic two-body forces
  could be handled, now in a set of coupled integral equations in
  two variables. That approach was based on Spline
  interpolations~\cite{glo82},
  which allowed to integrate analytically over the interpolated Faddeev
  amplitudes under the integral. In this manner in the
  three-nucleon system the realistic high precision nucleon-nucleon (NN)
  forces~\cite{AV18,CDBONN,NIJMI}
  together with three-nucleon forces (3NF) of various types~\cite{TM,uIX}
   can be handled in a
  fully reliable manner~\cite{glo96}. These studies are a basic foundation
  for testing nuclear forces and triggered a tremendously rich set
  of neutron-deuteron (nd) and proton-deuteron (pd)
  experiments all over the world. A representative set can be found 
 in~\cite{glo96,howell1988,stephan1989,rauprich1991,aninch93,sydow94,allet96,setze96,rohdjess98,abfalt98,sakai2000,bieber2000,cadman2001,ermisch2001,hatanaka2002,seki2004,ermisch2005,amir2007}. Especially the
  need of 3N forces in conjunction with current two-nucleon (2N) forces was
  firmly established~\cite{wit98,wit2001,wit2005,seki2005}.
  In recent years that interplay of 2N and 3N
  forces has been substantiated by the theoretical insights
  into nuclear forces gained through effective field theory for
  $\pi$'s and nucleons ($\Delta$'s) constrained by chiral
  symmetry~\cite{epel2006}.

The feasibility of  controlling the 3N continuum in the Faddeev
scheme opened the door to evaluate reliably the final state
(initial state) interaction in photon induced processes on $^3He$
(pd capture processes)~\cite{golak2005}. This provides important
insight into the electromagnetic nucleonic current operator, the
interplay with 3N forces and into properties of electromagnetic
nucleon form factors.

Despite that achieved technical status of controlling the integral
Faddeev kernel in the three-body continuum it would be desirable
to avoid those logarithmic singularities totally. A suggestion in
that direction was undertaken in~\cite{hub94} where, however, the
presence of the virtual state pole of the 2N t-matrix in the
state $^1S_0$  caused problems. It is  the purpose of this paper
to establish a definite solution of that long lasting technical
challenge with the moving logarithmic singularities  by
formulating the three-body Faddeev kernel in the continuum in such
a manner that only trivial poles occur in one variable, which can
be handled as easily as in the 2-body Lippmann Schwinger equation.

In section II we give a brief reminder of our standard approach
followed by the discussion of all possible alternative choices of
variables in the  intermediate state integral of the Faddeev
integral kernel. One of them sticks out which is free of all
technical difficulties and will be displayed in detail. In section
III we compare numerical results for  3N  scattering  in the old
and new approach using modern  high precision NN forces. 
The great simplification achieved with the novel approach shows up also 
in case of finite rank 2-body forces as displayed in section IV.  
 We conclude in section V.

\section{Different forms of the 3-body Faddeev equation in the continuum}

We use the notation detailed in~\cite{glo96}. The nucleon-deuteron
(Nd) breakup  amplitude is
\begin{eqnarray}
    < \vec p \vec q | U_0 | \Phi> = < \vec p \vec q | ( 1+P ) T|
    \Phi> ~,
    \label{breakup}
\end{eqnarray}
where $ \vec p $ and $ \vec q $ are standard Jacobi momenta, $ P $
the sum of a cyclical and anticyclical permutation of 3 particles
and $ | \Phi>$ the initial product state of a deuteron and a
momentum eigenstate of the projectile nucleon.

The amplitude $ T| \Phi> $ obeys our standard Faddeev type
    equation
\begin{eqnarray}
    T| \Phi> = t P | \Phi> + t P G_0 T | \Phi> ~,
    \label{T}
\end{eqnarray}
where $ t $ is the 2N off-shell t-operator and $ G_0$ the free
    3N propagator.
    Introducing the   momentum space 3N partial wave basis
     $ | pq \alpha>$ and projecting (\ref{T}) on these states we
    get
\begin{eqnarray}
\langle p q \alpha \vert T \vert \Phi \rangle = \langle p q \alpha
\vert  t P \vert \Phi \rangle
   + \langle p q \alpha \vert  t P G_0 T \vert \Phi \rangle ~.
\label{TPROJ}
\end{eqnarray}

The kernel part $\langle p q \alpha \vert  t P G_0 T \vert \Phi
\rangle$ can be evaluated using the completeness  of the states 
$|pq \alpha>$ as
\begin{eqnarray}
&&\langle p q \alpha \vert  t P G_0 T \vert \Phi \rangle
 = \sum_{\alpha~' \alpha''} \int p~'^2dp~'q~'^2dq~'p''^2dp''q''^2dq''
\langle p q \alpha \vert  t \vert p~' q~' \alpha~' \rangle \cr
 && \times \langle p~' q~' \alpha' \vert P \vert p'' q'' \alpha'' \rangle
{\frac{ \langle p'' q'' \alpha''  \vert T \vert \Phi \rangle} { { E + i \epsilon -\frac{1}{m}( p''^2 + \frac{3}{4} q''^2)  }} }  ~.
 \label{PWFD}
\end{eqnarray}

Here we use the Balian-Ber\'ezin~\cite{balian,keister2003} 
approach to calculate 
 the permutation
matrix element $\langle p~' q~' \alpha' \vert P \vert p'' q''
\alpha'' \rangle$ (see Appendix for details) 
\begin{eqnarray}
\langle p  q  \alpha \vert  P  \vert p' q'  \alpha' \rangle &=&
\int_{-1}^{1} dx { \frac {\delta(p-\pi_1)}  { p^{2} }  } ~
{ \frac {\delta(p~ '-\pi_2)}  { p~ '^{2} }  } ~ 
G_{\alpha \alpha'}^{BB} (q, q', x)~ .
\label{eqBBperm}
\end{eqnarray}

Moreover the 2-body t-matrix conserves the spectator momentum
 $q $ and all discrete quantum numbers except the orbital angular
 momentum $l$:
\begin{eqnarray}
  < pq \alpha| t | p~ ' q~ ' \alpha~ '> &=& \frac{\delta( q - q~
 ')}{q^2}t_{l_{\alpha}l_{\bar \alpha}}^{s_{\alpha} j_{\alpha}
 t_{\alpha}}( p p~ '; E(q) = E-\frac{3}{4m} q^2)  \cr
 &&  \delta_{s_{\alpha} s_{\alpha~ '}} \delta_{j_{\alpha} j_{\alpha~'}}
\delta_{t_{\alpha}
t_{\alpha~'}}\delta_{\lambda_{\alpha}\lambda_{\alpha~'}}
\delta_{I_{\alpha} I_{\alpha~'}} ~.
\end{eqnarray}

Finally we  extract the deuteron pole in the channels $ | \alpha> = |
\alpha_d> $ which contain the 2-body $^3S_1-^3D_1$ states.
 Thus we define
\begin{eqnarray}
t_{l_{\alpha} l_{\bar \alpha}}^{s_{\alpha} j_{\alpha} t_{\alpha}}
(p,p~';E(q)) \equiv \frac{\hat t_{l_{\alpha} l_{\bar
\alpha}}^{s_{\alpha} j_{\alpha} t_{\alpha}} (p,p~';E(q))}{E +i
\epsilon  - \frac{3}{4m} q^2 - E_d}
\end{eqnarray}
for the deuteron quantum numbers $s_{\alpha} = j_{\alpha }= 1,
 t_{\alpha }=0, l_{\alpha}, l_{\bar \alpha}=0,2$ and keep $t$ as
it is otherwise. That pole property obviously carries over to the
$ T$-amplitude and we define just for the $ | \alpha> = |
\alpha_d> $ channels
\begin{eqnarray}
\langle p q \alpha \vert T \vert \Phi \rangle = \frac{\langle p q
\alpha \vert  \hat  T \vert \Phi \rangle}{E +i \epsilon  -
\frac{3}{4m} q^2 - E_d} ~.
\end{eqnarray}

Using all that the coupled set (\ref{TPROJ}) is  for $\alpha \ne
\alpha_d$
\begin{eqnarray}
&&\langle p q \alpha \vert  T \vert \Phi \rangle  = \langle p
q\alpha \vert  t P  \vert \Phi \rangle +
 \sum_{l_{\bar \alpha}} \sum_{ \alpha''} 
\int p~'^2dp~'p''^2 dp'' q''^2 dq''\cr
 &&  \int_{-1}^{1} dx ~ t_{l_{\alpha} l_{\bar \alpha}}^{s_{\alpha}
j_{\alpha} t_{\alpha}} (p,p~';E(q)) 
   G^{BB}_{\bar \alpha \alpha~ ''}(q,q'',x) 
\frac{\delta( p~'- \pi_1)}{{p~'}^2}
  \frac{\delta( p''- \pi_2) }{{p''}^2} \cr
&&   (  \delta_{\alpha'' \alpha_d''}\frac{ \langle p'' q'' \alpha'' \vert
\hat T \vert \Phi \rangle}{E +i \epsilon  - \frac{3}{4m} q''^2 - E_d}
   + \bar \delta_{\alpha'' \alpha_d''} \langle p'' q'' \alpha''
\vert  T \vert \Phi \rangle) 
   \frac{1}{ E + i \epsilon -
\frac{1}{m}( p''^2 + \frac {3} {4} q''^2)  } ~,
\label{eqa}
\end{eqnarray}
where $ \bar \delta_{\alpha'' \alpha_d''} = 1 - \delta_{\alpha''
\alpha_d''}$, and for $ \alpha = \alpha_d $
\begin{eqnarray}
&&\langle p q \alpha_d \vert  \hat T \vert \Phi \rangle  = \langle
p q\alpha_d \vert  \hat t P  \vert \Phi \rangle +
 \sum_{l_{\bar \alpha}} \sum_{ \alpha''} 
\int p~'^2dp~'p''^2 dp'' q''^2 dq''\cr
 &&  \int_{-1}^{1} dx ~ \hat t_{l_{\alpha}
l_{\bar \alpha}}^{s_{\alpha} j_{\alpha} t_{\alpha}} (p,p~';E(q))
   G^{BB}_{\bar \alpha \alpha~ ''}(q,q'',x) 
\frac{\delta( p~'- \pi_1)}{{p~'}^2}
  \frac{\delta( p''- \pi_2) }{{p''}^2} \cr
  &&   (  \delta_{\alpha'' \alpha_d''}\frac{ \langle p'' q'' \alpha''
\vert \hat T \vert \Phi \rangle}{E +i \epsilon  - \frac{3}{4m} q''^2 - E_d}
   + \bar \delta_{\alpha'' \alpha_d''} \langle p'' q'' \alpha''
\vert  T \vert \Phi \rangle)   \frac{1}{ E + i \epsilon -
\frac{1}{m}( p''^2 + \frac {3} {4} q''^2)  } ~.
\label{eqb}
\end{eqnarray}

The shifted $ p~ '$ and $p~ ''$ arguments are
\begin{eqnarray}
 \pi_1  & = & \sqrt{q''^2 + \frac{1}{4} q^2 + q q''x} ~, \cr
 \pi_2 & = & \sqrt{q^2 + \frac{1}{4} q''^2 + q q''x} ~.
 \label{eq4}
\end{eqnarray}

We are left in Eqs. (\ref{eqa}) and (\ref{eqb})  with four
integrations over $p~', p'', q''$, and $x$, where any two of them
can be performed analytically using the two $\delta$-functions.
Thus  there are six possibilities, which we regard now in turn and
we discuss the advantages or disadvantages to use them. Apparently
it is sufficient to discuss only the kernel parts and moreover
just one, say in (\ref{eqa}), which we shall denote just by "the
kernel".

\subsection{Analytical integration over $ p~ ' $ and $ p'' $}
\label{sub1}

The resulting form of "the kernel"  is
\begin{eqnarray}
&&\langle p q \alpha \vert  t P G_0 T \vert \Phi \rangle =
  \sum_{l_{\bar \alpha}} \sum_{ \alpha''}  
\int q''^2dq'' \int_{-1}^{1} dx
 ~ t_{l_{\alpha} l_{\bar \alpha}}^{s_{\alpha} j_{\alpha}
t_{\alpha}} (p,\pi_1;E(q)) \cr
 &&    G^{BB}_{\bar \alpha \alpha~ ''}(q,q'',x) 
(  \delta_{\alpha'' \alpha_d''}\frac{ \langle \pi_2 q'' \alpha'' \vert
\hat T \vert \Phi \rangle}{E +i \epsilon  - \frac{3}{4m} q''^2 - E_d}
   + \bar \delta_{\alpha'' \alpha_d''} \langle \pi_2 q'' \alpha'' \vert  T
\vert \Phi \rangle)\cr
 &&   \frac{1}{ E + i \epsilon - \frac{1}{m}( q^2 + q''^2  + q q'' x )
 } ~.
 \label{kI}
\end{eqnarray}

This is our standard approach~\cite{wit86,glo96}. For $ q \le
q_{max}\equiv \sqrt{\frac{4}{3}mE}$ the  integration over $x$
leads to logarithmic singularities depending on $ q $ and $ q~ ''$
- the so called moving singularities. Nevertheless the advantage
is, that the complex $ q'' $-dependence of $  \langle \pi_2 q''
\alpha''
 \vert T \vert \Phi \rangle $ can be properly mapped out. That nontrivial
  dependence of the $ T $-amplitude arises from the property of the 2N t-matrix
  in the state $ ^1 S_0 $ and from the 3N breakup threshold
behavior~\cite{glo96}.

The deuteron pole at $ q'' = q_0 = \sqrt{\frac{4 m}{3}( E-E_d)} $
  can be  taken care of in the $q''$-integration using e.g. a 
subtraction method.

\subsection{Analytical integration over $ x $ and $ p~ ''$}
\label{sub2}

We rewrite the $\delta$-functions in (\ref{eqa}) and (\ref{eqb})
as follows
\begin{eqnarray}
&&  \delta( p~' - \pi_1)
 =   \frac{2 p~'}{ q q''} \delta( x - x_0) \Theta( 1 - | x_0|) ~,
 \label{pi1}
\end{eqnarray}
with
\begin{eqnarray}
x_0 = \frac{p~'^2 - 1/4 q^2 - q''^2}{q q''} = \frac{p''~^2 - 1/4 q''^2
- q^2}{q q''} ~,
\label{eq8}
\end{eqnarray}
and
\begin{eqnarray}
 \delta( p'' - \pi_2)= \delta( p~'' - \sqrt{p~ '^2 + \frac{3} {4}q^2 -
\frac {3} {4} q~''^2} ) \Theta( p~ '^2 + \frac{3} {4}q^2 - \frac
{3} {4} q~''^2) ~.
\label{eq7_1}
\end{eqnarray}

The resulting form of "the kernel" is
\begin{eqnarray}
&&\langle p q \alpha \vert  t P G_0 T \vert \Phi \rangle = 
\frac{2}{q} \sum_{l_{\bar \alpha}} \sum_{ \alpha''} 
 \int p~' d p~'  q'' d q''  t_{l_{\alpha} l_{\bar
\alpha}}^{s_{\alpha} j_{\alpha} t_{\alpha}} (p,p~';E(q)) 
    G^{BB}_{\bar \alpha \alpha~ ''}(q,q'',x_0) \cr &&
  \Theta( 1 - \vert \frac{p~'^2 - 1/4 q^2 - q~''^2}{q q~ ''}\vert)
\Theta( p~'^2 + \frac{3}{4} q^2 - \frac{3}{4} q~''^2)\cr
    &&(  \delta_{\alpha'' \alpha_d''}\frac{ \langle p'' q'' \alpha''
\vert \hat T \vert \Phi \rangle}{E +i \epsilon  - \frac{3}{4m} q''^2 - E_d}
   + \bar \delta_{\alpha'' \alpha_d''} \langle p'' q'' \alpha'' \vert  T
\vert \Phi \rangle)   \frac{1}{ E + i \epsilon -
\frac{1}{m}( {p~ '}^2 + \frac{3}{4} q^2 )  } ~,
 \label{kII}
\end{eqnarray}
 with
\begin{eqnarray}
p~ '' =\sqrt{ p~'^2 + \frac{3}{4} q^2 - \frac{3}{4} q~''^2} ~.
\label{eq.17}
\end{eqnarray}

\begin{figure}[htbp]
\includegraphics[scale=1.0]{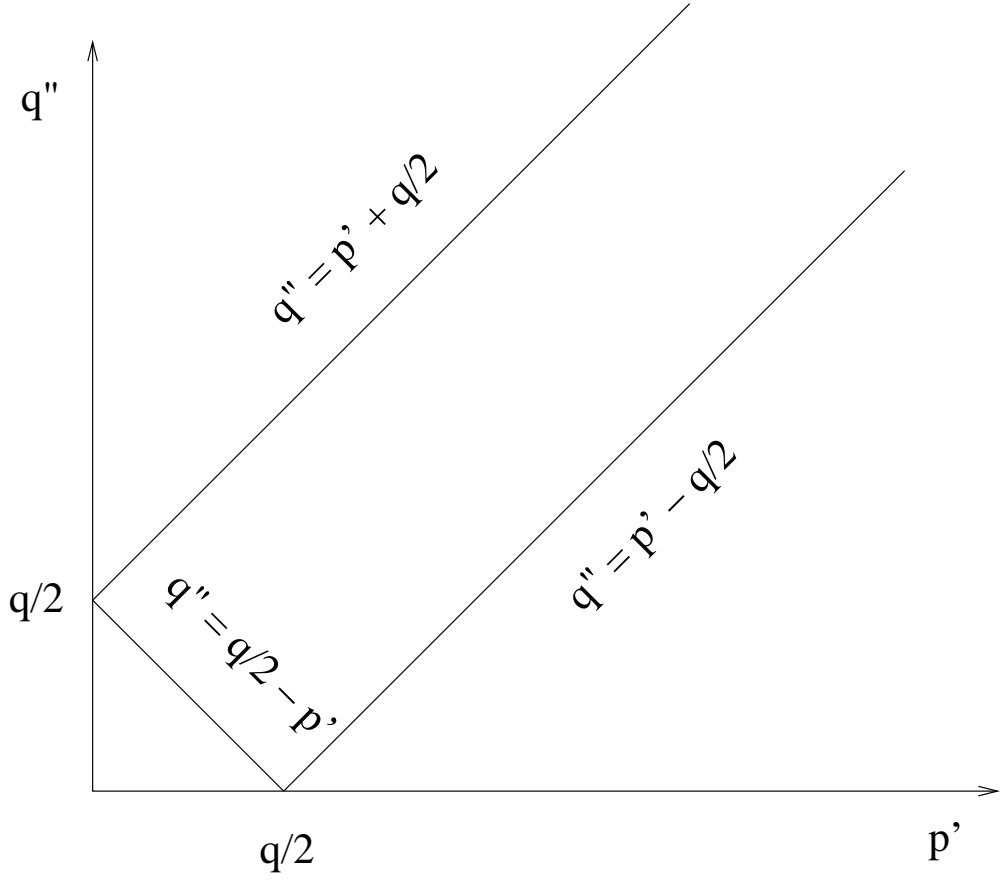}
\caption{The domain for the integrations over $p'$ and $q''$
(rectangular region contained between the three lines) in the case
\ref{sub2}.}
\label{fig1a}
\end{figure}

The two $\Theta$-functions define the domain D for the
integrations over $ p~'$ and $q''$ which is an open rectangular
region in the $ p~'-q''$ plane restricted by the straight lines $
q'' = \frac {q} {2} \pm p~' $ and $ q'' =  p~' - \frac{q} {2} $ as
displayed in Fig.~\ref{fig1a}.

The integrations over $p~'$ and $q''$ in (\ref{kII}) split now in
the following way
\begin{eqnarray}
&&\langle p q \alpha \vert  t P G_0 T \vert \Phi \rangle
  =
\frac{2}{q} \sum_{l_{\bar \alpha}} \sum_{ \alpha''}
 \int_0^{\infty} p~' d p~'t_{l_{\alpha} l_{\bar
\alpha}}^{s_{\alpha} j_{\alpha} t_{\alpha}} (p,p~';E(q))
 \frac{1}{ E + i \epsilon - \frac{1}{m}( {p~ '}^2 + \frac{3}{4} q^2 )
}\cr
& &  \int_{| q/2 - p~'|}^{q/2 + p~'} q'' d q''
    G^{BB}_{\bar \alpha \alpha~ ''}(q,q'',x_0) 
 (  \delta_{\alpha'' \alpha_d''}\frac{ \langle p'' q'' \alpha''
\vert \hat T \vert \Phi \rangle}{E +i \epsilon  - \frac{3}{4m} q''^2 -
E_d}
   + \bar \delta_{\alpha'' \alpha_d''} \langle p'' q'' \alpha'' \vert  T
\vert \Phi \rangle),
 \label{eq.18}
\end{eqnarray}
For the
channels $ \alpha'' $ different from $ \alpha_d'' $ only a simple pole in
the $ p'$ variable
occurs, positioned at $ p_0 = \sqrt{\frac{3}{4}( q_{max}^2 - q^2)}$ and
therefore of concern only for $ q
\le  q_{max}$. At $ q = q_{max}$ or $  p_0 = 0 $ there is no pole since the
$ q''$ integral
vanishes. 
For the channels $ \alpha'' = \alpha_d'' $ the $ q'' $-  integral
contains the deuteron pole at $ q'' = q_0 = \sqrt{\frac{4m}{3} ( E-E_d)}$. 
If $q_0$ does not coincide
with the limits of integration $ | \frac{q}{2} - p~'|$ 
and $\frac{q}{2} + p~'$ that integral generates a
smooth function of $ p~ '$. In case it coincides, however, 
 logarithmic singularities occur.
Their positions are
determined by
\begin{eqnarray}
q_0 &=& | \frac{q}{2} - p~ '| ~, \cr
q_0 &=& \frac{q}{2} + p~' ~.
\end{eqnarray}
We can avoid them in the following manner. The product of the free propagator 
 and
the deuteron pole term can be written as
\begin{eqnarray}
& & \frac{1}{ E + i \epsilon - \frac{1}{m}( {p~ '}^2 + \frac{3}{4} q^2)} 
~\frac{1}{E +i \epsilon  - \frac{3}{4m} q''^2 - E_d}\cr
&  = &  ( \frac{1}{ E + i \epsilon - \frac{1}{m}( {p~ '}^2 + \frac{3}{4} q^2 )}
 - \frac{1}{E +i \epsilon  - \frac{3}{4m} q''^2 - E_d} ) 
\frac{1}{ |E_d| + \frac{1}{m} p''^2} ~,
\end{eqnarray}
where we used (\ref{eq.17}). This provides separation of the free propagator 
 and the deuteron pole singularities. 
Thus for the $ \alpha''= \alpha_d'' $ channels alone one has the 
contribution to ``the kernel''
\begin{eqnarray}
 &&
\frac{2}{q} \sum_{l_{\bar \alpha}} \sum_{  \alpha_d''}
 \int_0^{\infty} p~' d p~'
t_{l_{\alpha} l_{\bar \alpha}}^{s_{\alpha} j_{\alpha}  
t_{\alpha}} (p,p~';E(q))
  \int_{| q/2 - p~'|}^{q/2 + p~'} q'' d q'' 
G^{BB}_{\bar \alpha \alpha_d''}(q,q'',x_0) \cr
& & [ \frac{1}{ E + i \epsilon - \frac{1}{m}( {p~ '}^2 + \frac{3}{4} q^2 )} ~
\frac{1}{|E_d| + \frac{1}{m} p''^2} 
< p'' q'' \alpha_d'' \vert \hat T \vert \Phi >\cr
&  - &  \frac{1}{E +i \epsilon  - \frac{3}{4m} q''^2 - E_d}  ~ 
\frac{1}{|E_d| + \frac{1}{m} p''^2}
<  p'' q'' \alpha_d'' \vert \hat T \vert \Phi>]\label{eq.22}
\end{eqnarray}

The first part has only the free propagator singularity in 
the $p~'$ variable like the part for
the channels $ \alpha''$ different from $ \alpha_d''$ in (\ref{eq.18})  
and in the second
part we change the order of integrations. Thus (\ref{eq.22}) yields the 
contribution to ``the kernel'' from $ \alpha'' = \alpha_d''$ channels
\begin{eqnarray}
 &&
\frac{2}{q} \sum_{l_{\bar \alpha}} \sum_{  \alpha_d''}
 \int_0^{\infty} p~' d p~' 
t_{l_{\alpha} l_{\bar \alpha}}^{s_{\alpha} j_{\alpha}  t_{\alpha}} (p,p~';E(q))
 \frac{1}{ E + i \epsilon - \frac{1}{m}( {p~ '}^2 + \frac{3}{4} q^2 )}\cr
& & \int_{| q/2 - p~'|}^{q/2 + p~'} q'' d q'' 
G^{BB}_{\bar \alpha \alpha_d''}(q,q'',x_0)
\frac{< p'' q'' \alpha_d'' \vert \hat T \vert \Phi >}
{|E_d| + \frac{1}{m} p''^2} \cr
& - &  \frac{2}{q} \sum_{l_{\bar \alpha}} \sum_{ \alpha_d''}
\int_0^{\infty} dq'' q''
\frac{1}{E +i \epsilon  - \frac{3}{4m} q''^2 - E_d} \cr
& &  \int_{| q/2 - q''|}^{q/2 + q''} p~' d p~' 
t_{l_{\alpha} l_{\bar \alpha}}^{s_{\alpha}
j_{\alpha}  t_{\alpha}} (p,p~';E(q))
 G^{BB}_{\bar \alpha \alpha_d''}(q,q'',x_0)
\frac{< p'' q'' \alpha_d'' \vert \hat T \vert \Phi>}{|E_d| + \frac{1}{m} p''^2} ~. 
\end{eqnarray}

Now in the second part the $p~'$-integration yields a smooth function in 
$q''$ and the deuteron
 singularity in the $q''$-integral is a simple pole.

Altogether ``the kernel'' (\ref{eq.18}) is
\begin{eqnarray}
&& < p q \alpha \vert  t P G_0 T \vert \Phi > = 
\frac{2}{q} \sum_{l_{\bar \alpha}} \sum_{ \alpha''}
 \int_0^{\infty} p~' d p~'
t_{l_{\alpha} l_{\bar \alpha}}^{s_{\alpha} j_{\alpha}  
t_{\alpha}} (p,p~';E(q))
 \frac{1}{ E + i \epsilon - \frac{1}{m}( {p~ '}^2 + \frac{3}{4} q^2 )} \cr
&& \int_{| q/2 - p~'|}^{q/2 + p~'} q'' d q'' 
G^{BB}_{\bar \alpha \alpha''}(q,q'',x_0) 
 (\bar \delta_{\alpha'' \alpha_d''} 
< p'' q'' \alpha'' | T| \Phi>  + \delta_{\alpha'' \alpha_d''}
\frac{< p'' q'' \alpha_d'' \vert \hat T \vert \Phi >}{|E_d| + \frac{1}{m} p''^2} ) 
 \cr
& - &  \frac{2}{q} \sum_{l_{\bar \alpha}} 
\sum_{ \alpha_d''}\int_0^{\infty} dq'' q''
\frac{1}{E +i \epsilon  - \frac{3}{4m} q''^2 - E_d} \cr
& &  \int_{| q/2 - q''|}^{q/2 + q''} p~' d p~' 
t_{l_{\alpha} l_{\bar \alpha}}^{s_{\alpha}
j_{\alpha}  t_{\alpha}} (p,p~';E(q))
 G^{BB}_{\bar \alpha \alpha_d''}(q,q'',x_0) 
\frac{< p'' q'' \alpha_d'' \vert \hat T \vert \Phi>}{|E_d| + \frac{1}{m} p''^2}  ~.
\label{eq.24}
\end{eqnarray}

It is the best form out of the six possible ones 
for the numerical performance.
 Also the complex q-dependence of the $T$-amplitude near $q''=q_{max}$ 
in the channels
 $\alpha$ containg the $^1S_0$ two-body component can be properly
mapped out like in case
 \ref{sub1}.

\subsection{Analytical integration over $ x $ and $p~ '$}
\label{sub3}

We choose one $\delta$-function in (\ref{eqa}) as in (\ref{pi1}) and the other
  one as
\begin{eqnarray}
&&  \delta( p'' - \pi_2) =   \frac{p~ ''}{  p~'} \delta( p~' -
\sqrt{p''^2 + \frac{3} {4}q''^2 - \frac {3} {4} q^2}~ ) \Theta( p''^2
+ \frac{3} {4}q''^2 - \frac {3} {4} q^2) ~.
\label{eq7}
\end{eqnarray}

This leads to "the kernel"
\begin{eqnarray}
&&\langle p q \alpha \vert  t P G_0 T \vert \Phi \rangle =
\frac{2} {q} \sum_{l_{\bar \alpha}} \sum_{ \alpha''} 
\int p'' dp'' \int q'' dq''
 ~   t_{l_{\alpha} l_{\bar \alpha}}^{s_{\alpha} j_{\alpha}
t_{\alpha}} (p,p~';E(q)) 
   G^{BB}_{\bar \alpha \alpha~ ''}(q,q'',x_0)  \cr
&&\Theta( qq'' - \vert p''^2 - \frac{1} {4}q''^2 -  q^2 \vert
) \Theta( p''^2+ \frac{3} {4}q''^2 - \frac {3} {4} q^2) \cr
&&(  \delta_{\alpha'' \alpha_d''}\frac{ \langle p'' q'' \alpha'' \vert
\hat T \vert \Phi \rangle}{E +i \epsilon  - \frac{3}{4m} q''^2 - E_d}
   + \bar \delta_{\alpha'' \alpha_d''} \langle p'' q'' \alpha''
\vert  T \vert \Phi \rangle)   \frac{1}{E + 
i \epsilon - \frac{1}{m} (p''^2 + \frac{3}{4} q''^2 )
   } ~,
 \label{kIII}
\end{eqnarray}
with
\begin{eqnarray}
p~' = \sqrt{p''^2 +3/4q''^2-3/4q^2} ~.
 \label{eq10}
\end{eqnarray}

The two $\Theta$-functions define the domain D for the
integrations over $ p''$ and $q''$ which is an open rectangular
region in the $ p''-q''$ plane restricted by straight lines $ q'' = 2
( q \pm p'') $ and $ q'' = 2 p'' - 2 q$ and which leads to the
integrals

\begin{eqnarray}
&&\langle p q \alpha \vert  t P G_0 T \vert \Phi \rangle =
\frac{2} {q} \sum_{l_{\bar \alpha}} \sum_{ \alpha''}
\int_0^{\infty} dp'' p'' \int_{| 2q - 2p''|}^{2q + 2q''} dq'' q'' 
 t_{l_{\alpha} l_{\bar \alpha}}^{s_{\alpha} j_{\alpha} t_{\alpha}} (p,p~';E(q))
   G^{BB}_{\bar \alpha \alpha~ ''}(q,q'',x_0)  \cr
&&(  \delta_{\alpha'' \alpha_d''}\frac{ \langle p'' q'' \alpha'' \vert
\hat T \vert \Phi \rangle}{E +i \epsilon  - \frac{3}{4m} q''^2 - E_d}
   + \bar \delta_{\alpha'' \alpha_d''} \langle p'' q'' \alpha''
\vert  T \vert \Phi \rangle)   \frac{1}{E + i \epsilon - \frac{1}{m} 
(p''^2 + \frac{3}{4} q''^2 )} ~,
 \label{kIIIa}
\end{eqnarray}

One can proceed analogously like in \ref{sub2} and would arrive also at 
a form free of logarithmic
singularities. The form \ref{sub2} might appear, however, more 
favorable since each propagator appears
only with one integration variable.

\subsection{Analytical integration in $ q''$ and $ p~ '$}
\label{sub4}

We keep the first $\delta$-function  in (\ref{eqa}) as it is and
    rewrite the second one as
\begin{eqnarray}
&&  \delta( p'' - \pi_2) = \frac {2 p''} {  \sqrt{p''^2 - q^2(1 -
x^2) } } \delta( q'' - (-2 q x + 2 \sqrt{p''^2
-q^2(1-x^2) } ) \Theta( p''^2 - q^2) ~.
 \label{eq17}
\end{eqnarray}

The resulting form of "the kernel"  is
\begin{eqnarray}
&&\langle p q \alpha \vert  t P G_0 T \vert \Phi \rangle =
 2 \sum_{l_{\bar \alpha}} \sum_{ \alpha''}  
 \int p'' dp'' \Theta(p''^2 -q^2) 
 \int_{-1}^{1} dx  
 ~ \frac { t_{l_{\alpha} l_{\bar \alpha}}^{s_{\alpha} j_{\alpha}
t_{\alpha}} (p,\pi_1;E(q))} { \sqrt{p''^2 - q^2(1-x^2) }} \cr
&&    G^{BB}_{\bar \alpha \alpha~ ''}(q,q'',x)  
(\delta_{\alpha'' \alpha_d''}\frac{ \langle p'' q'' \alpha''
\vert \hat T \vert \Phi \rangle}{E +i \epsilon  - \frac{3}{4m} q''^2 - E_d}
   + \bar \delta_{\alpha'' \alpha_d''} \langle p'' q'' \alpha''
\vert  T \vert \Phi \rangle)\cr & & \frac{1}{ E + i \epsilon -
p''^2 - 3 ( \sqrt{p''^2 - q^2(1-x^2)} -qx )^2 }  ~,
\label{kIV}
\end{eqnarray}
with
\begin{eqnarray}
q'' = 2(\sqrt{p''^2 - q^2(1-x^2)} - qx) ~.
\label{qbis}
\end{eqnarray}

This form appears quite complicated in the pole structure of the
free propagator as well as for the deuteron pole and the $ ^1S_0$
virtual state pole singularity close to $q''=q_{max}$ and we do not
consider it further.

\subsection{Analytical integration over $ q''$ and $ p''$}
\label{sub5}

We rewrite the first $\delta$-function in (\ref{eqa}) as
\begin{eqnarray}
 \delta( p~ ' - \pi_1) = \frac{p' \delta( q'' - (
\sqrt{ p~ '^2 - \frac{1}{4} q^2 ( 1- x^2)} -
\frac{qx}{2}))}{\sqrt{p~ '^2 - \frac{1}{4} q^2 ( 1-
    x^2)}}\Theta( {p~ '}^2 - 1/4 q^2) ~,
\end{eqnarray}
and keep the second as
\begin{eqnarray}
    \delta( p~ '' -\sqrt{( q - 1/2 q~ '')^2 + q q~ '' ( 1+x)}) ~.
\end{eqnarray}

Thus $ p~ '' $ is always greater equal zero. It is zero for
$x=-1$ and $ p~ ' = 3/2 q$.

"The kernel"  is
\begin{eqnarray}
&&\langle p q \alpha \vert  t P G_0 T \vert \Phi \rangle 
 = \sum_{l_{\bar \alpha}} \sum_{ \alpha''} 
\int dp~'\Theta( {p~ '}^2 - 1/4 q^2)\cr
 &&  \int_{-1}^{1} dx
 ~ t_{l_{\alpha} l_{\bar \alpha}}^{s_{\alpha} j_{\alpha}
t_{\alpha}} (p,p~';E(q))  
\frac{p' q''^2}  {\sqrt{p'~^2 - \frac{1} {4} q^2(1-x^2)} } 
G^{BB}_{\bar \alpha \alpha~ ''}(q,q'',x)   \cr
  &&(\delta_{\alpha'' \alpha_d''}\frac{ \langle p'' q'' \alpha'' 
\vert \hat T \vert \Phi \rangle}{E +i
\epsilon  - \frac{3}{4m} q''^2 - E_d} 
     + \bar \delta_{\alpha'' \alpha_d''} \langle  p'' q'' 
   \alpha'' \vert  T \vert \Phi \rangle) 
   \frac{1} { E + i \epsilon - \frac{1}{m}( {p~ '}^2
+ \frac {3}  {4} q^2) } ~,\label{kV}
\end{eqnarray}
with
\begin{eqnarray}
p~ '' = \sqrt{( q - \frac{1}{2} q~ '')^2 + q q~ '' ( 1+x)}
~,
\end{eqnarray}
and
\begin{eqnarray}
q~ '' =\sqrt{ p~ '^2 - \frac{1}{4} q^2 ( 1- x^2)} - \frac{qx}{2}
~.
\end{eqnarray}

Though the free propagator singularity is again a simple pole in $p~'$ 
a two-fold integration in the $ T$-amplitude is required.
This appears to be a disadvantage (though surmountable) against
the case \ref{sub2}. However, in the deuteron pole both 
integration variables $p'$ and $x$ occur, which is quite unfavorable.

\subsection{Analytical integration over $x$ and $ q~ ''$}
\label{sub6}

The first $\delta$-function in (\ref{eqa}) is taken as in
  (\ref{pi1}) and the second one as
\begin{eqnarray}
   \delta(p~ '' - \sqrt{q^2 + 1/4 q''^2 + q q''x}) &=& 
   \frac{4 p~ ''}{3 q~''}\delta ( q~ '' - \sqrt{4/3 ( {p~ '}^2 - p''^2)
+ q^2}) \cr
&& \Theta(4/3 ( {p~ '}^2 - p''^2) + q^2) ~.
\end{eqnarray}

It results in "the kernel"
\begin{eqnarray}
&&\langle p q \alpha \vert  t P G_0 T \vert \Phi \rangle 
 = \frac{8}{3q} \sum_{l_{\bar \alpha}} \sum_{ \alpha''}
 \int dp~' p~'\int dp'' p''
\Theta( 4/3 ( {p~ '}^2 - p''^2) + q^2) \Theta( 1 - | x_0|) \cr
 &&  t_{l_{\alpha}
l_{\bar \alpha}}^{s_{\alpha} j_{\alpha} t_{\alpha}}  (p,p~';E(q)) 
   G^{BB}_{\bar \alpha \alpha~ ''}(q,q'',x_0)  \cr
&&(\delta_{\alpha'' \alpha_d''}\frac{ \langle p''
q~ '' \alpha'' \vert \hat T \vert \Phi \rangle}
   {E - E_d +i \epsilon  - \frac{1}{m}( {p~ '}^2 - p''^2) - \frac{3}{4m} q^2}
   + \bar \delta_{\alpha'' \alpha_d''} \langle p'' q~ '' 
 \alpha'' \vert  T \vert \Phi \rangle)\cr && \frac{1} { E + i
\epsilon - \frac{1}{m}( {p~ '}^2 + \frac{3}{4} q^2) } ~,
\end{eqnarray}
with
\begin{eqnarray}
q~ '' =  \sqrt{\frac{4}{3} ( {p~ '}^2 - p''^2) + q^2}
~.
\end{eqnarray}

Though the free propagator singularity is just a simple pole in
one variable in the deuteron pole both integration variables $ p~
'$ and $ p~ ''$ occur, which is less favorable.

We conclude that case \ref{sub2} is clearly the most favorable
choice and we compare in the next section  results for different
3N observables choosing our standard approach, case \ref{sub1},
and that new one, case \ref{sub2}.

\section{Comparison of the new (\ref{sub2}) and standard (\ref{sub1}) 
approaches}

\begin{figure}[htbp]
\includegraphics[scale=0.96]{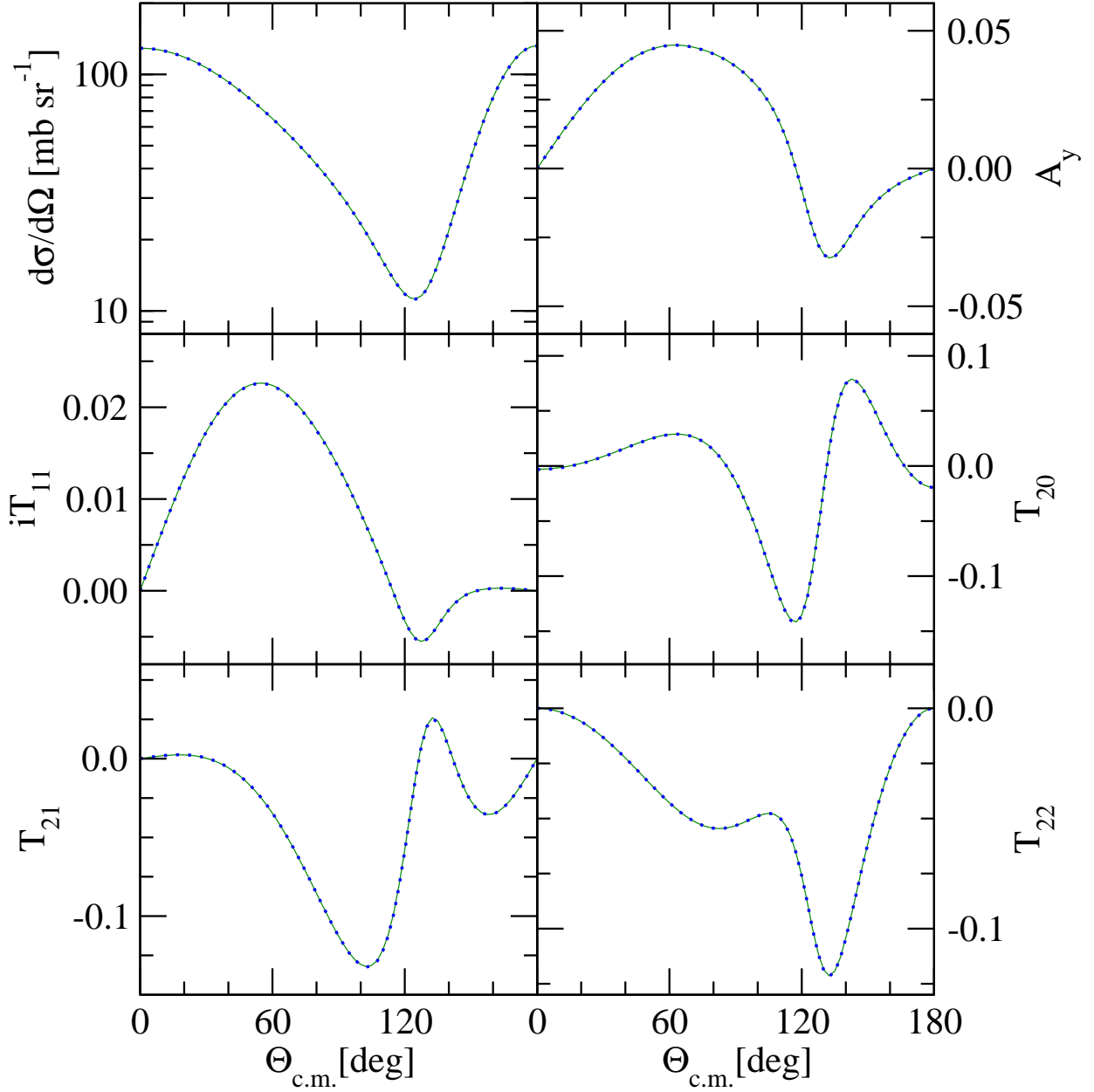}
\caption{
The nd elastic scattering angular distribution and analyzing powers
at an incoming neutron lab. energy $E^n_{lab}=13$~MeV.
The solid line is the CD~Bonn potential prediction using our standard
approach of handling the logarithmic singularities.  The
dotted line is  obtained with the new approach without  
logarithmic singularities.
 The 2-nucleon states are kept up to $ j_{max }=1$ }.
\label{fig1}
\end{figure}

\begin{figure}[htbp]
\includegraphics[scale=1.0]{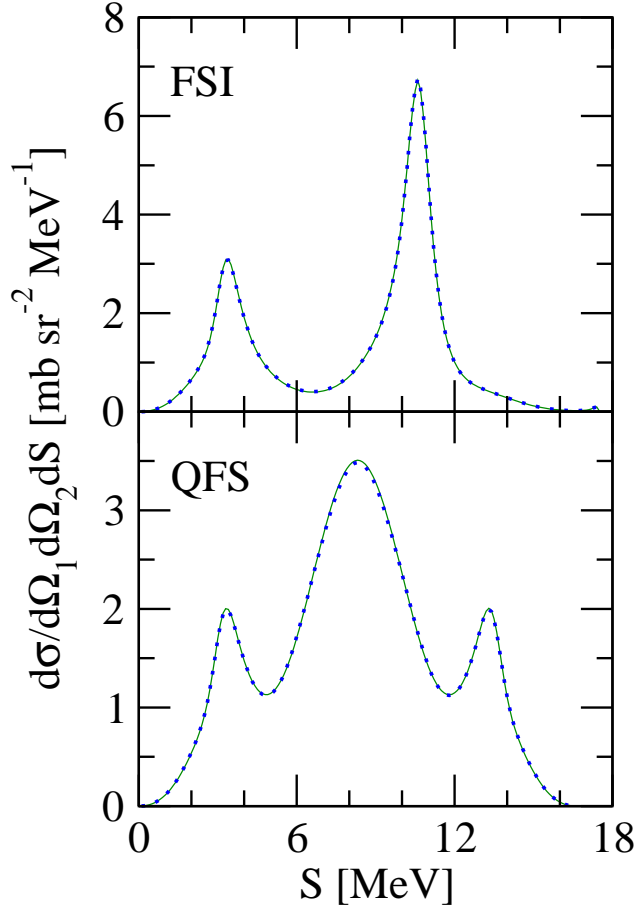}
\caption{
 Cross sections for the exclusive d(n,nn)p breakup
 at an incoming neutron lab. energy $E^n_{lab}=13$~MeV.
Curves as in Fig 2.  The upper configuration is the
final-state-interaction (FSI) geometry with the polar angles of the 
detected neutrons
$\theta_1=39^o$, $\theta_2=62.5^o$ and the azimuthal angle $\phi_{12}=180^o$.
The lower configuration is the quasi-free-scattering (QFS)
 geometry with the polar angles of detected neutrons
$\theta_1=\theta_2=39^o$ and the azimuthal angle $\phi_{12}=180^o$.
}
\label{fig2}
\end{figure}

Since the complicated  singularity pattern in the old approach exists only for  $q'' \le q_{max}$ 
we applied the new approach only there  and 
kept the old one for  $q'' > q_{max}$,  where only a simple 
deuteron pole is present. 
We used the CD~Bonn~\cite{CDBONN} potential 
restricted to act in the two-nucleon 
partial wave states 
with total angular momentum $j \le 1$.  
In Fig.~\ref{fig1} we show the resulting nd elastic 
scattering angular distributions for the cross section and various  
analyzing powers 
at an incoming neutron lab. energy $E^n_{lab} = 13$~MeV.  
 The agreement obtained with the two approaches is very good. The cross 
sections at the same 
energy for two geometries of the Nd breakup
 are shown in Fig.~\ref{fig2}. Again the agreement is very good.

\section{Finite rank forces}

Using the choice \ref{sub2} simplifies the treatment of the 
3N continuum in case 
of finite rank 2-body
forces also very significantly since there are no longer 
logarithmic singularities.
  For the sake of a simple notation 
we keep only s-waves and
restrict the 2-body force to act only in the states $^1S_0$ and $^3S_1$. 
This leads to two coupled
equations for the two amplitudes $T_k( pq)$ where $k=1~ (k=2)$ goes 
with $^1S_0~ (^3S_1)$, respectively.
Choosing the kernel of the type (\ref{eq.24}) one obtains explicitely
\begin{eqnarray}
T_1(pq) &= &  T_1^0(pq) + \frac{2}{q} \int_0^{\infty} dp~' p~'t_1( p,p~';E(q)) 
\frac{1}{E + i \epsilon - \frac{1}{m}( p~'^2 +\frac{3}{4} q^2)}\cr
& &  \int_{| q/2- p~ '|}^{q/2 + p~'} d q'' q''  ( G_{11} T_1( p'' q'') 
+ G_{12} \frac{\hat T_2 ( p''q'')} {E_d +\frac{1}{m}p''^2})\cr
& - &  \frac{2}{q}\int_0^{\infty} d q'' q''
\frac{1}{E +  i \epsilon - \frac{3}{4m} q''^2 - E_d} G_{12}
\int_{| q/2 - q''|}^{q/2 + q''} dp~' p~'t_1( p,p~';E(q)) 
\frac{\hat T_2 ( p''q'')} {E_d +\frac{1}{m}p''^2}\\
\hat T_2 ( pq)&  
= &  \hat T_2^0 ( pq) + \frac{2}{q} 
\int_0^{\infty} dp~' p~'\hat t_2( p,p~';E(q))
\frac{1}{E + i \epsilon - \frac{1}{m}( p~'^2 +\frac{3}{4} q^2)}\cr
& & \int_{| q/2- p~ '|}^{q/2 + p~'} d q'' q'' ( G_{21} T_1( p'' q'') + 
G_{22} \frac{\hat T_2 ( p''q'')} {E_d
+\frac{1}{m}p''^2})\cr
& - &  \frac{2}{q}\int_0^{\infty} d q'' q''
 \frac{1}{E +  i \epsilon - \frac{3}{4m} q''^2 - E_d} G_{22}
 \int_{| q/2 - q''|}^{q/2 + q''} dp~' p~' 
\hat t_2( p,p~';E(q))\frac{\hat T_2 ( p''q'')}
 {E_d +\frac{1}{m}p''^2}
\end{eqnarray}

with 
\begin{eqnarray}
p'' = \sqrt{p~'^2 + \frac{3}{4} q^2 - \frac{3}{4} q''^2}\label{p2} ~.
\end{eqnarray}

Now we assume the finite rank forms
\begin{eqnarray}
t_1 ( p p~'; E(q))&  = &  g_1(p) \tau_1( E(q)) g_1( p~')\label{sep.t1}~,\cr
t_2 ( p p~';E(q)) & = &  g_2 (p) \frac{\hat \tau_2( E(q))}{ E + 
i \epsilon - \frac{3}{4m} q^2 - E_d} g_2(p~') ~,
\label{sep.3}
\end{eqnarray}
and obtain
\begin{eqnarray}
T_1(pq) &= &  T_1^0(pq) +  g_1(p) \tau_1( E(q))[\frac{2}{q} 
\int_0^{\infty} dp~' p~'g_1( p~')
\frac{1}{E + i \epsilon - \frac{1}{m}( p~'^2 +\frac{3}{4} q^2)}\cr
& &  \int_{| q/2- p~ '|}^{q/2 + p~'} d q'' q''  ( G_{11} T_1( p'' q'')
+ G_{12} \frac{\hat T_2 ( p''q'')} {E_d +\frac{1}{m}p''^2})\cr
& - &  \frac{2}{q}\int_0^{\infty} d q'' q''
\frac{1}{E +  i \epsilon - \frac{3}{4m} q''^2 - E_d} G_{12}
\int_{| q/2 - q''|}^{q/2 + q''} dp~' p~'g_1( p~')
 \frac{\hat T_2 ( p''q'')} {E_d +\frac{1}{m}p''^2}]\\
\hat T_2 ( pq)&  = &  \hat T_2^0 ( pq) + g_2 (p) 
\hat \tau_2( E(q))[\frac{2}{q}
 \int_0^{\infty} dp~' p~'\hat g_2( p~') 
\frac{1}{E + i \epsilon - \frac{1}{m}( p~'^2 +\frac{3}{4} q^2)}\cr
& & \int_{| q/2- p~ '|}^{q/2 + p~'} d q'' q'' ( 
G_{21} T_1( p'' q'') + G_{22}  \frac{\hat T_2 ( p''q'')} {E_d +\frac{1}{m}p''^2})\cr
& - &  \frac{2}{q}\int_0^{\infty} d q'' q'' \frac{1}{E +  i \epsilon - \frac{3}{4m} q''^2 - E_d} G_{22}
 \int_{| q/2 - q''|}^{q/2 + q''} dp~' p~' g_2( p~'
)\frac{\hat T_2 ( p''q'')} {E_d +\frac{1}{m}p''^2}]
\end{eqnarray}
where going through the same steps it results
\begin{eqnarray}
T_1^0(pq) & = &  N g_1(p) \tau_1(E(q)) F_1^0(q) ~,\cr
\hat T_2^0(pq)&  = &  N g_2(p) \hat \tau_2(E(q)) \hat F_2^0(q) ~.
\label{sep.7}
\end{eqnarray}
The normalisation factor $N$ provides the dependence 
on spin and isospin quantum numbers ($m_d$, $m_0$ are spin projections 
of the initial deuteron and nucleon, respectively, and $\nu_0$  
the nucleon's  isospin projection),
on  the
initial momentum $ q_0$ and the deuteron normalisation 
factor $ N_d$  defined as
 $ \phi_d( p) = N_d \frac{g_2(p)}{E_d - \frac{p^2}{m}} $:
\begin{eqnarray}
N = \frac{1}{\sqrt{4 \pi}} \delta_{M_T, \nu_0}   ( 1 
\frac{1}{2}  \frac{1}{2}, m_d, m_0, M)
\frac{N_d}{q_0} ~.
\label{sep.8}
\end{eqnarray}
Further $ F_1^0(q) $ and $ \hat F_2^0(q) $ are given as
\begin{eqnarray}
F_1^0(q) & = &  G_{12} \frac{2}{q} \int_{|q_0 - q/2|}^{q_0 
+ q/2} dp~' p~' g_1(p~')
  \frac{g_2 ( \sqrt{p~'^2 + \frac{3}{4} q^2 - \frac{3}{4} q_0^2})}
{E_d -  \frac{p~'^2 + \frac{3}{4} q^2 - \frac{3}{4} q_0^2}{m}} ~, \cr
\hat F_2^0(q) & = &  G_{22} \frac{2}{q} 
\int_{|q_0 - q/2|}^{q_0 + q/2} dp~' p~' g_2(p~')
  \frac{g_2 ( \sqrt{p~'^2 + \frac{3}{4} q^2 - \frac{3}{4} q_0^2})}
{E_d -  \frac{p~'^2 + \frac{3}{4} q^2 - \frac{3}{4} q_0^2}{m}} ~.
\label{sep.10}
\end{eqnarray}
It follows the structures
\begin{eqnarray}
T_1(pq) &=& g_1(p) \tau_1( E(q)) F_1(q) ~, \cr
\hat T_2(pq)  &=& g_2 (p) \hat \tau_2( E(q)) \hat F_2(q) ~,
\label{sep.12}
\end{eqnarray}
and therefore one obtains the two coupled one-dimensional equations 
inserting explicitely the
integration limits

\begin{eqnarray}
F_1(q) &  = &  F_1^0(q) + \frac{2}{q} \int_0^{\infty} dp~'p~'g_1(p~')
\frac{1} {E + i \epsilon - \frac{1}{m}( p~'^2 + \frac{3}{4} q^2)}\cr
& &  \int_{|q/2 - p~'|}^{q/2 + p~'} d q'' q''  ( 
G_{11} g_1(p'') \tau_1( E(q'')) F_1(q'')
+ G_{12} \frac{g_2 (p'') 
\hat \tau_2( E(q''))\hat F_2 ( q'')} {E_d +\frac{1}{m}p''^2})\cr
& - &  \frac{2}{q}\int_0^{\infty} d q'' q''
\frac{G_{12}}{E +  i \epsilon - \frac{3}{4m} q''^2 - E_d}
\int_{|q/2 - q''|}^{q/2 + q''} dp~' p~'g_1( p~') 
\frac{g_2 (p'') \hat \tau_2( E(q'')) \hat F_2(q'')} {E_d +\frac{1}{m}p''^2}]\\
\hat F_2 ( pq)&  = &  \hat F_2^0 ( q) + \frac{2}{q}
 \int_0^{\infty} dp~' p~'\hat g_2( p~')
 \frac{1}{E + i \epsilon - \frac{1}{m}( p~'^2 +\frac{3}{4} q^2)}\cr
& & \int_{|q/2 - p~'|}^{q/2 + p~'} d q'' q''
 ( G_{21} g_1(p'') \tau_1( E(q'')) F_1(q'') + G_{22}  
\frac{g_2 (p'') \hat \tau_2( E(q''))
 \hat F_2(q'')} {E_d +\frac{1}{m}p''^2})\cr
& - &  \frac{2}{q}\int_0^{\infty} d q'' q'' 
\frac{G_{22}}{E +  i \epsilon - \frac{3}{4m} q''^2 - E_d} 
 \int_{|q/2 - q''|}^{q/2 + q''} dp~' p~' g_2( p~')
\frac{g_2 (p'') \hat \tau_2( E(q'')) \hat F_2(q'')} {E_d +\frac{1}{m}p''^2}]
\end{eqnarray}

It remains to provide the factors
\begin{eqnarray}
G_{11} &=& \frac{\sqrt{2}}{8} = G_{22}  ~, \cr
G_{12} &=&  - \frac{3 \sqrt{2}}{8} = G_{21} ~.
\label{sep.16}
\end{eqnarray}

Note for $ E = \frac{3}{4m} q^2 $ there is no singularity at $ p~'=0 $ 
since the $ q''$-integral vanishes
at $ p~ '=0 $. Also there occur only simple poles, which can be treated  by
subtraction.  Using for instance Spline interpolation for $ F_1(q'') $ 
and $ \hat F_2(q'')$ based
on a set of grid points  the two integrals can be trivially 
performed and one obtains
 a low dimensional  inhomogeneous algebraic set of equations.

\section{Summary and conclusions}

Starting from a partial wave decomposed form of the 3N Faddeev equation for 
a breakup amplitude
we discussed the six possible choices of integrating over internal angular 
and momentum
variables in the integral kernel. While our standard approach integrates 
over the moving
logarithmic singularities along the real momentum axis we found a new 
one, which totally avoids
that technical obstacle. It is very simple. The free 3N propagator 
singularity appears as a pole
in a single variable, which can be taken care of trivially (like in the 
2-body Lippmann-Schwinger equation). 
The deuteron pole singularity is again a simple pole 
and moreover the two
poles are cleanly separated in two different integration variables.

The other four  choices for internal integration variables turned out to 
be less favorable.

We numerically  compared the two approaches evaluating some nd elastic 
and breakup 
observables and found very good agreement. This should open the door to 
handle the 3N
 continuum as simply as solving the 2-body Lippmann-Schwinger equation -  
though of course 
some more variables occur. It will also simplify the application to 
electromagnetic
 processes in the 3N system, where initial and final state interactions have to
 be treated properly.

We also draw  attention to the Balian-Ber\'ezin method, which was proposed 
long time ago,
 and which deserves much more attention than it received up to now.

The inclusion of 3N forces do not change the singularity structure of 
the kernel and can be equally well treated in the new 
approach but is left to a
forthcomig study.

\section*{Acknowledgments}
This work was  partially supported by the Helmholtz Association
through funds provided to the virtual institute ``Spin and strong
QCD''(VH-VI-231) and by the  Polish Committee for Scientific
Research. The numerical calculations were performed on the IBM
Regatta p690+ of the NIC in J\"ulich, Germany.

\appendix 

\section{Permutation operator}
\label{a2}

In view of using the Balian-Ber\'ezin
method~\cite{balian,keister2003} for evaluation of the permutation
matrix element $\langle p~' q~' \alpha' \vert P \vert p'' q''
\alpha'' \rangle$ it is adequate to change from $jJ$ coupling to
$LS$ coupling
\begin{eqnarray}
|pq \alpha> = \sum_{\beta} | pq \beta> < \beta | \alpha> ~,
\end{eqnarray}
where
\begin{eqnarray}
&& < \beta | \alpha> = < ( l \lambda ) L ( s \frac{1}{2} ) S ( LS) J 
(t \frac{1}{2})T | (ls)
 j ( \lambda \frac{1}{2} ) I ( j I ) J (t \frac{1}{2})T > 
 = \sqrt{\hat j \hat I \hat L \hat S} 
\left\{ \begin{matrix}
 l & s & j \cr
                \lambda & \frac{1}{2} & I \cr
                L & S & J \cr
\end{matrix}
                \right\} ~.
\end{eqnarray}

The permutation matrix  element is~\cite{glo83}
\begin{eqnarray}
&& \langle p~ ' q~ ' \beta~ '\vert P \vert p~ '' q~ '' \beta~
''\rangle = 2 \delta_{S~ ' S~ ''}   
\delta_{T~ ' T~ ''}   
\delta_{L~ ' L~ ''} \delta_{\mu~ ' \mu~ ''} \cr
&& (-)^{s~ ''} \sqrt{\hat s~ ' \hat s~ ''}
 \left\{ \begin{matrix}
 \frac{1}{2} & \frac{1}{2} & s~ ' \cr
 \frac{1}{2} & S~ ' & s~ '' \cr
\end{matrix}
                \right\} 
(-)^{t~ ''} \sqrt{\hat t~ ' \hat t~ ''}
 \left\{ \begin{matrix}
 \frac{1}{2} & \frac{1}{2} & t~ ' \cr
 \frac{1}{2} & T~ ' & t~ '' \cr
\end{matrix}
                \right\}  \cr
 && < p~ ' q~ ' ( l~ ' \lambda~ ') L~ ' \mu| P_{12}P_{23} | p~ '' q~
'' ( l~ '' \lambda~ '') L~ ' \mu> ~.
\label{PERM1}
\end{eqnarray}

Since the momentum space matrix element in (\ref{PERM1}) is
independent of $ \mu$ one can put
\begin{eqnarray}
&& \langle p~ ' q~ ' ( l~ ' \lambda~ ') L~ '\vert P_{12}P_{23}
\vert p~ '' q~ '' ( l~ '' \lambda~ '') L~ '\rangle\cr &=&
\frac{1}{2 L~ ' + 1} \sum_{\mu} \langle p~ ' q~ ' ( l~ ' \lambda~
') L~ '\mu \vert P_{12}P_{23} \vert p~ '' q~ '' ( l~ '' \lambda~
'') L~ '\mu \rangle ~.
\end{eqnarray}

Further we use
\begin{eqnarray}
< \vec p~ ' \vec q~ '| pq ( l \lambda) L \mu> = \frac{\delta( p~'
- p) }{p^2}\frac{\delta( q~ ' - q) }{q^2} Y_{l \lambda}^{L
\mu}(\hat p~ ' \hat q~ ') ~,
\end{eqnarray}
and the linear relations among the Jacobi momenta of different
types and obtain
\begin{eqnarray}
&& \langle p~ ' q~ ' ( l~ ' \lambda~ ') L~ '\vert P_{12}P_{23}
\vert p~ '' q~ '' ( l~ '' \lambda~ '') L~ '\rangle\cr
 &=& \frac{1}{2 L~ ' +
1} \sum_{\mu~} \int d \hat p~ ' d \hat q~ ' d \hat p~ '' d \hat q~
'' Y_{l~ '\lambda~ '}^{L~ ' \mu^{  *}}( \hat p~ ' \hat q~
')Y_{l~''\lambda~ ''}^{L~ ' \mu~  }( \hat p~ '' \hat q~ '') \cr
 && \delta ( \vec p~ ' - 1/2 \vec q~ ' - \vec q~ '')
  \delta ( \vec p~ '' + \vec q~ ' + 1/2 \vec q~ '') ~.
\end{eqnarray}

Here enters the idea of Balian-Ber\'ezin~\cite{balian}. The sum over
the products of spherical harmonics is a scalar and depends only
on the scalar products among the four unit vectors. Due to the two
$\delta$-functions all those scalar products are fixed by the
magnitudes of the four momenta $p', q', p''$, and $q''$. Thus we introduce
\begin{eqnarray}
  X( p~ ' q~ ' p~ '' q~ '') \equiv\frac{1}{2 L~ ' + 1} \sum_{\mu~}
Y_{l~ '\lambda~ '}^{L~ ' \mu^{  *}}( \hat p~ ' \hat q~
')Y_{l~''\lambda~ ''}^{L~ ' \mu }( \hat p~ '' \hat q~ '') ~.
\end{eqnarray}

Consequently $ X $ can be taken out of the integral and one
obtains
\begin{eqnarray}
&& \langle p~ ' q~ ' ( l~ ' \lambda~ ') L~ '\vert P_{12}P_{23}
\vert p~ '' q~ '' ( l~ '' \lambda~ '') L~ '\rangle\  =  \cr 
& &  8 \pi^2 \int_{-1}^1 dx ~ X( p~ ' q~
' p~ '' q~ '') \frac{ \delta( p~ '
- \vert 1/2 \vec q~ ' + \vec q~ '' \vert}{{p~'}^2}
 \frac{ \delta( p~ '' - \vert \vec q~ ' + 1/2 \vec q~ '' \vert
 }{{p''}^2} ~,
\end{eqnarray}
where $ x = \hat q~ ' \cdot \hat q $.

Finally we combine geometrical factors and write the permutation matrix 
element in the form
\begin{eqnarray}
\langle p  q  \alpha \vert ~ P ~ \vert p'  q'  \alpha' \rangle &=&
\int_{-1}^{1} dx { \frac {\delta(p-\pi_1)} { p^{2} }  } ~
{ \frac {\delta(p~ '-\pi_2)}  { p~ '^{2} }  } ~ 
G_{\alpha \alpha'}^{BB} (q, q', x) ,
\label{eq20}
\end{eqnarray}
with
\begin{eqnarray}
\pi_1 &=& \sqrt{q'^2 +\frac {1} {4}q^2 + qq'x}  \cr
\pi_2 &=& \sqrt{q^2 + \frac {1} {4}q'^2 + qq'x} ~,
\label{eq21a}
\end{eqnarray}
and
\begin{eqnarray}
&&G^{BB}_{\alpha \alpha'}(q q' x) = (4\pi)^{3/2}~ \delta_{TT'}
\delta_{M_T M_{T'}}
\sqrt{\hat j \hat I \hat s \hat t \hat {\lambda} }
\sqrt{\hat j~ ' \hat I~ ' \hat s~ ' \hat t~ '  } 
(-1)^{s~ '+ t~ '} 
 \left\{
\begin{array}{ccc}
1/2 & 1/2 & t \cr
1/2 &  T  & t~ ' \cr
\end{array}
 \right\} 
\cr
&&\sum_{LS} \hat S  
\left\{ \begin{matrix}
 l & s & j \cr
                \lambda & \frac{1}{2} & I \cr
                L & S & J \cr
\end{matrix}
                \right\} 
\left\{ \begin{matrix}
 l~ ' & s~ ' & j~ ' \cr
                \lambda~ ' & \frac{1}{2} & I~ ' \cr
                L & S & J \cr
\end{matrix}
                \right\} 
 \left\{
\begin{array}{ccc}
1/2 & 1/2 & s \cr
1/2 &  S  & s~ ' \cr
\end{array}
  \right\}
\sum_{m_l m_{l~ '} m_{\lambda~ '}} ( l \lambda L,m_l 0 m_l) 
(l~ ' \lambda~ ' L, m_{l~'} m_{\lambda~ '} m_l) \cr
&& (-)^{m_l} Y_{l~ -m_l} (\hat p) Y_{l~ '~ m_{l~ '}} (\hat p~ ') 
Y_{\lambda~ ' m_{\lambda~ '}} (\hat q~ ') ~.
\label{eq23}
\end{eqnarray}

We use our standard notation $\hat l \equiv 2l+1$. It is assumed that
the z-axis is along $\vec q$ and the momentum $\vec q~'$ lies in the x-z
plane. That  leads to the following components of the $\vec q$, 
$\vec q~'$, $\vec p$, and $\vec p~'$ vectors
\begin{eqnarray}
\vec q &=& [0, 0, q] ~, \cr
\vec q~' &=& [q~ '\sqrt{1-x^2}, 0, q~ 'x] ~, \cr
\vec p &=& [q~ '\sqrt{1-x^2}, 0, q~ 'x + \frac {1} {2}q] ~, \cr
\vec p~' &=& [-\frac {1} {2}q~ '\sqrt{1-x^2}, 0, 
-q- \frac {1} {2}q'x] ~.
\label{eq24}
\end{eqnarray}

\clearpage

\end{document}